\documentclass[aps,prc,twocolumn,superscriptaddress]{revtex4-2}

\usepackage{graphicx}
\usepackage{braket}
\usepackage{bm}
\usepackage{color}
\usepackage{ulem}
\usepackage{amsmath}

\newcommand{\nucl}[2]{{}^{#1}\mathrm{#2}}
\newcommand{\ACa}{\alpha + \nucl{44}{Ca}}
\newcommand{\md}{\mathrm{d}}
\newcommand{\nn}{\nonumber}

\begin{document}


\title{Unexpectedly enhanced $\bm \alpha$-particle preformation in
$^{\bm{48}}$Ti probed by the $\bm{(p,p\alpha)}$ reaction} 

\author{Yasutaka~Taniguchi}
\email[]{taniguchi-y@di.kagawa-nct.ac.jp}
\affiliation{Department of Information Engineering, National Institute of Technology
(KOSEN), Kagawa College, Mitoyo, Kagawa 769-1192, Japan}
\affiliation{Research Center for Nuclear Physics (RCNP), Osaka University, Ibaraki
567-0047, Japan}
\author{Kazuki~Yoshida}
\affiliation{Advanced Science Research Center, Japan Atomic Energy Agency, Tokai,
Ibaraki 319-1195, Japan}
\affiliation{Research Center for Nuclear Physics (RCNP), Osaka University, Ibaraki
567-0047, Japan}
\author{Yohei~Chiba}
\affiliation{Department of Physics, Osaka City University, Osaka 558-8585, Japan}
\affiliation{Nambu Yoichiro Institute of Theoretical and Experimental Physics (NITEP),
Osaka City University, Osaka 558-8585, Japan}
\affiliation{Research Center for Nuclear Physics (RCNP), Osaka University, Ibaraki
567-0047, Japan}
\author{Yoshiko~Kanada-En'yo}
\affiliation{Department of Physics, Kyoto University, Kyoto 606-8502, Japan}
\affiliation{Research Center for Nuclear Physics (RCNP), Osaka University, Ibaraki
567-0047, Japan}
\author{Masaaki~Kimura}
 \affiliation{Department of Physics, Hokkaido University, Sapporo 060-0810, Japan}
 \affiliation{Nuclear Reaction Data Centre, Hokkaido University, Sapporo 060-0810,
 Japan}
 \affiliation{Research Center for Nuclear Physics (RCNP), Osaka University, Ibaraki
567-0047, Japan}
\author{Kazuyuki~Ogata}
\affiliation{Research Center for Nuclear Physics (RCNP), Osaka University, Ibaraki
567-0047, Japan}\affiliation{Department of Physics, Osaka City University, Osaka 558-8585, Japan}
\affiliation{Nambu Yoichiro Institute of Theoretical and Experimental Physics (NITEP),
Osaka City University, Osaka 558-8585, Japan}

\date{\today}

\begin{abstract}
 The formation of $\alpha$ particle on nuclear surface has been a fundamental problem since the 
 early age of nuclear physics. It strongly affects the $\alpha$ decay lifetime of heavy and
 superheavy elements, level scheme of light nuclei, and the synthesis of the elements in
 stars. However, the $\alpha$-particle formation in medium-mass nuclei has been poorly known
 despite its importance. Here, based on the $^{48}{\rm Ti}(p,p\alpha)^{44}{\rm Ca}$
 reaction analysis, we report that the $\alpha$-particle formation in a medium-mass nucleus
 $^{48}{\rm Ti}$ is much stronger than that expected from a mean-field approximation, and the
 estimated average distance between $\alpha$ particle and the residue is as large as 4.5~fm. This new result poses a challenge of describing four nucleon correlations by microscopic
 nuclear models. 
\end{abstract}


\maketitle


{\it ---Introduction.}
Since Gamow explained the $\alpha$ decay as the quantum tunneling of $\alpha$ particle out 
of an atomic nucleus \cite{Gamow1928}, the formation of $\alpha$ particle in nuclei has been a 
fundamental subject for understanding the structure and decay of nuclei
\cite{Mang1964,Jackson1977,Lovas1998,Qi2019}. It has been considered that $\alpha$ particles are
formed at a low-density nuclear surface with a certain probability, which is called the preformation
factor or the $\alpha$-particle preformation probability. It determines the lifetime of heavy and superheavy
nuclei, and its empirical values have often been estimated from the $\alpha$ decay lifetime. For
instance, the very short lifetime of $^{108}{\rm Xe}$ and $^{104}{\rm Te}$ were recently measured 
\cite{Auranen2018,Xiao2019}, and the enhancement of the $\alpha$-particle preformation probability beyond
proton-rich nucleus $^{100}{\rm Sn}$ has been discussed \cite{Clark2020,Mercier2020,Yang2020}.

It is also well known that the $\alpha$-particle preformation manifests itself in light nuclei as 
 $\alpha$ clustering \cite{Ikeda1968,Wildermuth1977} and is closely related to the synthesis of 
elements in stars \cite{Adelberger1998,Descouvemont2010}. Because it exhibits the unique excitation
spectra, $\alpha$ clustering has been identified in many light nuclei
\cite{fujiwara1980,Bijker2020}. Compared to heavy or light mass nuclei, the $\alpha$-particle preformation in
medium-mass nuclei has been poorly known. Generally, it is believed that $\alpha$-particle preformation is
hindered in medium-mass nuclei because of the largely negative $\alpha$-decay $Q$-values. The deep
binding energies of these nuclei also lead to the dominance of the mean-field dynamics over the
four nucleon correlation preventing $\alpha$-particle formation. However, such hindrance of $\alpha$-particle preformation has never been quantitatively confirmed by experiment due to the lack of reliable
 measure for the $\alpha$-particle preformation. 

The proton-induced $\alpha$-knockout reaction $(p, p\alpha)$ 
has been expected as the sensitive probe for the $\alpha$-particle preformation
\cite{Roos1977,Nadasen1980,Carey1984,Yoshimura1998,Mabiala2009}. Due to the strong absorption
effect, the $\alpha$ particle kicked by the projectile proton cannot get out from the interior of
the target nucleus. Consequently, the reaction is only sensitive to the $\alpha$ particles formed
on the surface of the target nucleus. Several experiments have been conducted to measure the
$\alpha$-particle preformation probability in light-medium mass nuclei. Carey {\it et al.}
reported a systematic measurement of the $(p, p\alpha)$ reactions with various target nuclei from
$^{16}{\rm O}$ to $^{66}{\rm Zn}$ \cite{Carey1984}. However, due to the lack of quantitative
analysis, the absolute value of the $\alpha$-particle preformation probabilities deduced from the
cross sections have large uncertainty. 

Recently, it has been shown that the distorted wave impulse approximation (DWIA) with reliable
optical potentials realizes an accurate description of the $(p,p\alpha)$
reaction~\cite{Yoshida2019}. Taking well-known light-mass $\alpha$
clustered nucleus 
$^{20}{\rm Ne}$ as an example, it was demonstrated that the $\alpha$-particle preformation probability is
quantitatively evaluated. The new analysis showed that the $\alpha$-particle preformation 
probability of $^{20}{\rm Ne}$ is smaller than that estimated by Carey {\it et al.} by a factor of
two. Among the nuclei studied by Carey {\it et al.}, $^{48}{\rm Ti}$ is the only one except 
for $^{20}{\rm Ne}$, for which the optical potentials between a proton, $\alpha$ particle, the residue ($\nucl{44}{Ca}$), and the target nucleus ($^{48}{\rm Ti}$) have already been known
accurately \cite{Delbar1978,Hama1990,Cooper1993}. Furthermore, the residue $\nucl{44}{Ca}$ is a magic
stable nucleus as an inert 
core, and hence, the enhancement of the $\alpha$-particle preformation can be expected. Therefore, the DWIA
analysis of the $^{48}{\rm Ti}(p,p\alpha)^{44}{\rm Ca}$ reaction must shed new insight into the
$\alpha$-particle preformation in medium-mass nuclei.  

{\it ---DWIA framework.} The DWIA framework \cite{Yoshida2016,Wakasa2017,Yoshida2018, 
Yoshida2019} has been adopted to describe the $^{48}{\rm Ti}(p,p\alpha)^{44}{\rm Ca}$
reaction. Within the factorization approximation, the triple differential cross section 
is given as,
\begin{align}
 \frac{\md^3\sigma}{\md T_p \md\Omega_p \md\Omega_\alpha} =C_0F_{\rm kin}
 \frac{\md\sigma_{p\alpha}}{\md\Omega_{p\alpha}}
 \left| \bar{T} \right |^2,
\end{align}
where $T_p$, $\Omega_p$ and $\Omega_\alpha$ are the kinetic energy of the emitted proton, the solid
angles of the proton and $\alpha$ particles, respectively. $C_0F_{\rm kin}$ is the kinematical
factor, and $\md\sigma_{p\alpha}/\md\Omega_{p\alpha}$ is the $p$-$\alpha$ differential cross section 
at the $p$-$\alpha$ relative momentum of the $(p,p\alpha)$ reaction kinematics. The
detail of this approximation is given in Refs.~\cite{Yoshida2016,Wakasa2017}, and the validity
of this approximation has been tested and confirmed \cite{Yoshida2016}. The reduced transition matrix element $\bar{T}$
is defined as, 
\begin{align}
\bar{T} &=
\int d^3R\, F(\bm{R}) y(R) Y_{00}(\hat{\bm{R}}), \label{eq:tmat}\\
 F(\bm{R}) &= \chi_{p}^{*(-)}(\bm{R}) \chi_{\alpha}^{*(-)}(\bm{R})
 \chi_{p}^{(+)}(\bm{R}) e^{-i\bm{k}_0\cdot\bm{R}/12},
\label{eq_F}
\end{align}
where $\bm{k}_0$ is the momentum of the incident proton. Equation (\ref{eq:tmat}) shows the
sensitivity of the cross section to the $\alpha$-particle preformation because it depends on the probability
amplitude of the $\alpha$-particle preformation $y(R)$. The other ingredients of the analysis are the optical
potentials for the $p$-$^{48}{\rm Ti}$, $p$-$^{44}{\rm Ca}$, and 
$\alpha$-$^{44}{\rm Ca}$ scattering, which are used to describe 
the distorted waves $\chi_{p}^{(\pm)}(\bm R)$ and $\chi_{\alpha}^{(-)}(\bm R)$; the superscripts $(+)$ and $(-)$ indicate outgoing and incoming boundary conditions, respectively.
It was shown that the use of the
accurate optical potentials is essential for the precise description of the cross sections and the
evaluation of $\alpha$-particle 
preformation. In the present work, the EDAD1 optical potential ~\cite{Hama1990,Cooper1993} with Dirac
phenomenology has been adopted to the $p$-$\nucl{48}{Ti}$ and $p$-$\nucl{44}{Ca}$ distorted
waves. This potential reproduces the proton-nucleus elastic scattering with various stable
targets from $^{12}{\rm C}$ to $^{208}{\rm Pb}$ in a wide energy range from 20 MeV to 1 GeV. For
the $\alpha$-$\nucl{44}{Ca}$ distorted wave, we applied the optical potential proposed by Delbar 
{\it et al.}~\cite{Delbar1978}, which reproduces the elastic differential cross sections from 24.1 to
100~MeV very accurately. All these optical potentials cover the required energy range for the 
analysis of the $^{48}{\rm Ti}(p,p\alpha)^{44}{\rm Ca}$ reaction.

{\it ---The $\alpha$-particle preformation probability.}
The probability amplitude for $\alpha$-particle preformation, called the reduced width amplitude (RWA),
is defined as, 
\begin{align}
 y(R) = \sqrt{\frac{48!}{4!~44!}}
 \Braket{\delta(r-R)\Phi_{\alpha}\Phi_{\rm Ca}
 Y_{00}(\hat r)|\Phi_{\rm Ti}}/R^2,\label{eq:rwa} 
\end{align}
where $\Phi_\alpha$, $\Phi_{\rm Ca}$, and $\Phi_{\rm Ti}$ denote the ground state wave functions of
the $\alpha$ particle, the residue ($\nucl{44}{Ca}$), and the target nucleus ($^{48}{\rm Ti}$), respectively. 
In this work, the $\alpha$ is assumed to have a $(0s)^4$ configuration, and the wave
functions of $^{44}{\rm Ca}$ and $^{48}{\rm Ti}$ are described by using the antisymmetrized molecular
dynamics (AMD) \cite{Kanada-Enyo2003,Kanada-Enyo2012,Kimura2016}. 
The parity-projected AMD wave function is given as, 
\begin{align}
 \Psi &= (1+P_x)/2\times\mathcal{A}\Set{\varphi_1\varphi_2...\varphi_A},\\
 \varphi_i&= \prod_{\sigma = x, y, z}
 \exp\Set{-\nu_\sigma\left(r_\sigma - Z_{i\sigma}\right)^2 }\nn\\
 &\times (\alpha_i \ket{\uparrow} + \beta_i \ket{\downarrow})\times (\ket{p}\mbox{ or }\ket{n}), 
\end{align}
where $P_x$ is the parity operator, $\mathcal{A}$ is the antisymmetrizer and $\varphi_i$ is the
nucleon wave packet. The centroid of a nucleon wave packet is a complex vector $\bm Z_i$,
in which the real (imaginary) part describes the mean position (momentum) of a nucleon. The parameters
of the model wave function are the centroids $\bm Z_i$, the spin directions $\alpha_i$ and $\beta_i$,
and the Gaussian widths $\nu_x$, $\nu_y$, and $\nu_z$. The wave function of $^{44}{\rm Ca}$ is
calculated within the mean-field approximation, i.e., the parameters are optimized to minimize the 
intrinsic energy $E=\braket{\Psi|H|\Psi}/\braket{\Psi|\Psi}$. Here, the Hamiltonian consists
of the nucleon kinetic energies, the effective nucleon-nucleon interaction, and the Coulomb
interaction. As an effective nucleon-nucleon interaction, we have used Gogny D1S density
functional~\cite{Berger1991} that reasonably reproduces the fundamental nuclear properties. After
the energy minimization, the $^{44}{\rm Ca}$ wave function is projected to $J^\pi=0^+$ to calculate
the RWA [Eq.~(\ref{eq:rwa})] using the Laplace expansion method \cite{Chiba2017}.
\begin{figure}[tbp]
 \centering
 \begin{tabular}{cc}
 \includegraphics[width=\hsize]{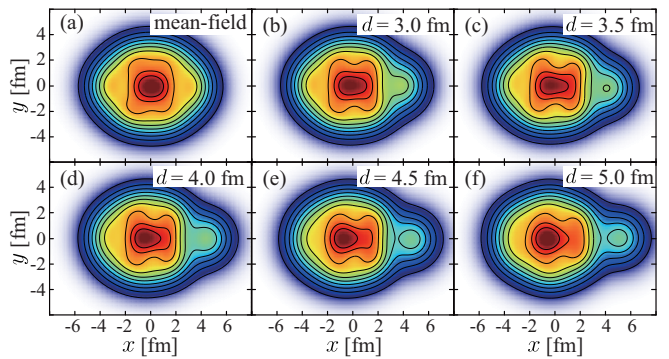}
 \end{tabular}
 \caption{Density distributions of the wave functions from which the RWAs have been calculated. (a)
 The mean-field solution. (b)--(f) The $\ACa$ system with various inter-nuclear distance $d$. }
 \label{fig:density} 
\end{figure}
\begin{figure}[tbp]
 \includegraphics[width=\hsize]{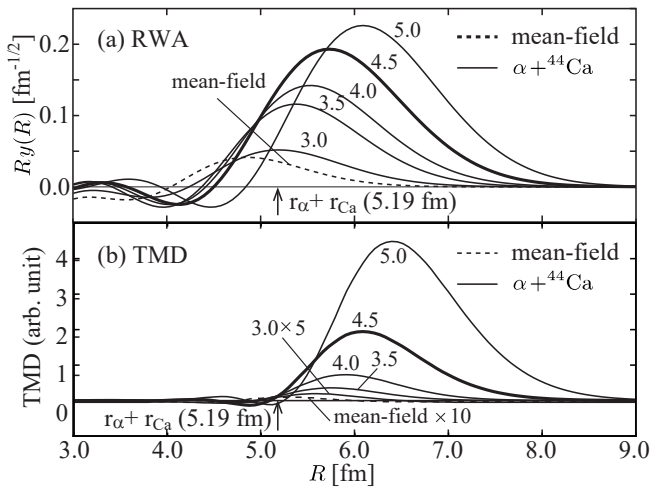}
 \caption{(a) The RWA calculated from the wave functions shown in Fig.~\ref{fig:density}.
 (b) The TMD of the $^{48}{\rm Ti}(p,p\alpha)^{44}{\rm Ca}$ reaction at $T_p=63$ MeV. The TMD
 obtained from the mean-field solution and the $d=3.0$~fm wave function are multiplied by a factor of
 10 and 5, respectively. The arrow indicates the sum of the charge radii of $\alpha$ and
 $^{44}{\rm Ca}$, which approximately corresponds to the nuclear surface.} \label{fig:rwa} 
\end{figure}

The wave function of $^{48}{\rm Ti}$ is also calculated in the same manner. The obtained wave
function, i.e., the mean-field solution for $^{48}{\rm Ti}$, is shown in Fig. \ref{fig:density}
(a). It has an almost spherical shape and does not clearly show the $\alpha$-particle preformation.
Indeed, the RWA calculated from this mean-field solution [Fig.~\ref{fig:rwa} (a)] has only a 
small peak at $R=4.8$~fm, and as discussed later, it is too small to reproduce the observed cross
section. Therefore, we artificially generate the test wave functions of $^{48}{\rm Ti}$ that
exhibit prominent $\alpha$-particle preformation. For this purpose, we introduce an approximate inter-nuclear distance $d$ \cite{Taniguchi2004, Taniguchi2020}, 
\begin{align}
 d=\left|\frac{1}{4}\sum_{i=1, ..., 4}{\rm Re}\bm Z_i - \frac{1}{44}\sum_{i=5, ..., 48} 
 {\rm Re}\bm Z_j \right|,\label{eq:dcnst}
\end{align}
where the first and second terms correspond to the center-of-mass of $\alpha$ and
$^{44}{\rm Ca}$, respectively. We perform the energy variation with the constraint on the value of
$d$ and obtain the wave functions which mimic the $\alpha$-particle preformation with
various inter-nuclear distance as shown in Fig.~\ref{fig:density} (b)--(f). The RWAs calculated from
these wave functions shown in Fig.~\ref{fig:rwa} (a) have prominent peaks that become higher and
move outward with the increase of $d$. Note that the RWAs are strongly suppressed in the 
interior of the residual nucleus ($R \lesssim 5$~fm) due to the Pauli principle. Consequently, the
peak position is not necessarily the same as the value of $d$.

\begin{figure}[tbp]
 \includegraphics[width=0.9\hsize]{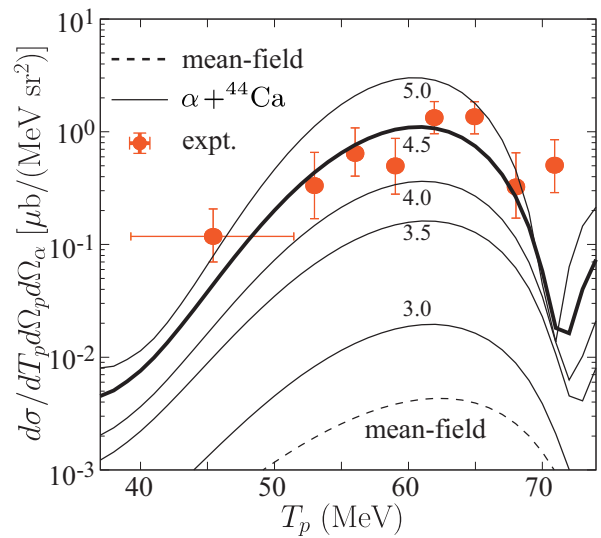}
 \caption{Triple differential cross section of the $^{48}{\rm Ti}(p,p\alpha)^{44}{\rm Ca}$ reaction
 obtained by the DWIA calculations using the RWAs shown in Fig.~\ref{fig:rwa} (a) compared with the
 experiment \cite{Carey1984}. The incident proton energy, the emitted angles of proton and $\alpha$ are
 set to $E_p=101.5$ MeV, $\theta_p=-70.0^\circ$ and $\theta_\alpha=45.0^\circ$, respectively. } \label{fig:cs} 
\end{figure}

\begin{figure}
 \centering
 \includegraphics[width=0.95\hsize]{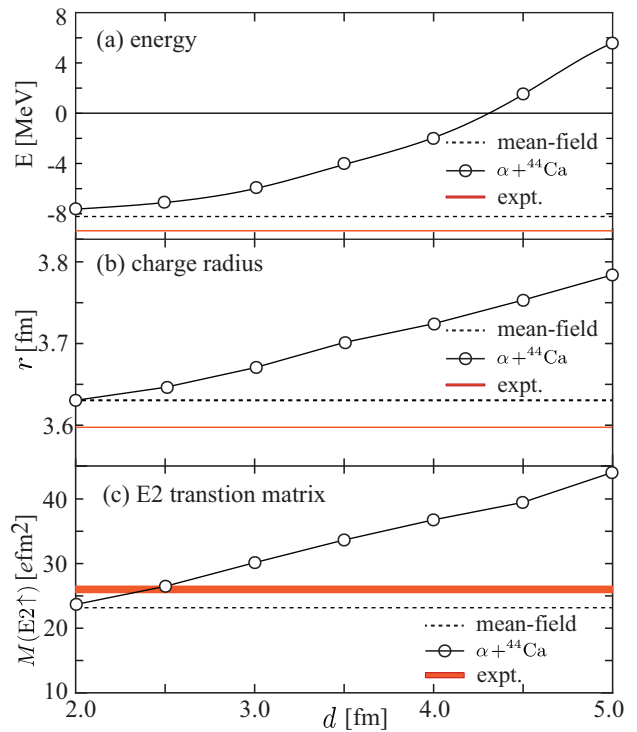}\\
 \caption{(a) Binding energy (b) charge radii, and (c) E2 transition matrix of $^{44}{\rm Ti}$
 calculated from the mean-field solution and $\ACa$ wave functions in comparison with the
 experimental data \cite{Huang2017,Angeli2013,Burrows2006}. The binding energy is given relative to
 the $\alpha$+$^{44}{\rm Ca}$ decay threshold. }
 \label{fig:radius}
\end{figure}

{\it ---Results and Discussions.}
Figure~\ref{fig:cs} shows the triple differential cross sections of the 
$\nucl{48}{Ti}(p, p\alpha)\nucl{44}{Ca}$ reaction obtained by the DWIA calculations using the RWAs
shown in Fig.~\ref{fig:rwa} (a). The cross sections are plotted as functions of the outgoing proton
energy. The incident proton energy, the emitted angles of the proton and $\alpha$ particle are
set to the same values as in the experiment by Carey {\it et al.}~\cite{Carey1984}. Unexpectedly, it
is found that the mean-field solution does not reproduce the observed cross section at all. It
underestimates the cross section in three orders of magnitude, which cannot be explained by the
uncertainty of the optical potentials used in DWIA analysis or the density functional (Gogny
D1S) used to calculate the mean-field solution. Consequently, we conclude that the $\alpha$
preformation probability is much larger than that described by the mean-field solution. 

To estimate the degree of $\alpha$-particle preformation, we have also tested the RWAs obtained from
the $\ACa$ wave functions with various inter-nuclear distances. Figure \ref{fig:cs} shows that these
RWAs 
yield much larger cross sections than the mean-field solution, and the cross section increases by
approximately one order of magnitude for every 1~fm increase of the inter-nuclear distance. It is
found that the RWA obtained from the $\ACa$ wave function with $d=4.5$~fm gives the most plausible
description of the observed cross section. The peripherality of the $(p,p\alpha)$ reaction is
confirmed from the real part of the transition matrix density (TMD) \cite{Wakasa2017} that is defined as,
\begin{align}
 \delta(R) = \bar{T}^*\int d\hat {\bm R}~R^2F(\bm R) y(R) Y_{00}(\hat {\bm R}).
\end{align}
Note that the integral of TMD over the distance is equal to the square of the transition matrix
$T$, and hence, $\delta(R)$ gives a hint at which distance $R$ the reaction takes places. As shown
in Fig. \ref{fig:rwa} (b), TMD is negligible in the interior region ($R \lesssim 5$~fm)
due to the strong absorption of an $\alpha$ particle and small RWA. It explains why the cross section
with the mean-field solution is smaller in order of magnitude than that with the $\ACa$ wave functions. We also
note that the peak position ($T_p\sim63$ MeV) and width of the cross section are approximately
determined by the kinematical condition (recoil-less condition for the residue $^{44}{\rm
Ca}$) and the momentum distribution of the RWA, respectively.

Although the $\ACa$ wave function with $d=4.5$~fm gives the best result for the
$^{48}{\rm Ti}(p,p\alpha)^{44}{\rm Ca}$ reaction, its validity should be verified from different
perspectives. Firstly, it must be noted that the binding energies of the $\ACa$ wave functions are
much smaller than that of the mean-field solution because of the artificial constraint imposed on
the inter-nuclear distance [Eq. (\ref{eq:dcnst})]. Figure \ref{fig:radius} (a) shows that the
binding energy of the $\ACa$ wave 
function rapidly decreases as the inter-nuclear distance increases. At $d=$4.5~fm, it underestimates
the experimental value \cite{Huang2017} by approximately 10 MeV and yields the positive $Q$-value of the
$\alpha$ decay, whereas the mean-field solution gives reasonable binding energy and $Q$-value. Panels
(b) and (c) show the charge radius and the reduced matrix elements for the 
E2 transition from the ground state to the $2^+_1$ state, respectively. As expected, 
both the charge radius and E2 transition matrix elements increase with the inter-nuclear
distance. Although the $\ACa$ wave function gives reasonable values at $d=2.0$--2.5~fm, it
overestimates the observed values \cite{Angeli2013,Burrows2006} at $d=4.5$~fm. In short, the $\ACa$
wave function can describe the $^{48}{\rm Ti}(p,p\alpha)^{44}{\rm Ca}$ reaction, but it fails to
reproduce the fundamental structural properties. On the contrary, the mean-field solution offers
a better description of the energy, radius, and E2 transition but fails in the $\alpha$ knockout
reaction. From these results, we can deduce that the ground state wave function should be an
admixture of the mean-field solution and the $\ACa$ type wave functions. The mean-field solution
should be the dominant component of the ground state due to its large binding energy, but the
contamination of the $\ACa$ wave function is indispensable to explain the observed large $\alpha$
knockout cross section. 

{\it ---Summary.} The $^{48}{\rm Ti}(p,p\alpha)^{44}{\rm Ca}$ reaction has been studied to
investigate the $\alpha$-particle preformation in a medium-mass nucleus $^{48}{\rm Ti}$. The DWIA analysis using
accurate optical potentials offers a reliable and
quantitative description of the $\alpha$-knockout reaction, and it has revealed that the $\alpha$-particle
preformation in $^{48}{\rm Ti}$ is unexpectedly enhanced. It has been shown that the mean-field solution
underestimates the cross section in orders of magnitude, and one must assume the $\ACa$ wave
function whose the inter-nuclear distance is as large as $d=4.5$~fm to reproduce the observed cross
section. However, the $\ACa$ wave function fails to explain other basic properties of
$^{48}{\rm Ti}$, which are reasonably described by the mean-field approximation. Hence, we conclude
that the ground state is an admixture of the mean-field and $\ACa$ configurations. This new insight
requests the systematic analysis of the $(p,p\alpha)$ reactions to reveal the universality of the
$\alpha$-particle preformation and poses a challenge to the microscopic nuclear models for describing
$\alpha$-particle preformation in medium-mass nuclei.

\acknowledgments
 This work was supported by the COREnet program at the RCNP, Osaka University, the Hattori Hokokai Foundation Grant-in-Aid for Technological and Engineering Research, and JSPS KAKENHI Grant Nos. JP16K05352, JP18H05407, JP18K03617, and JP20K14475.
Numerical calculations were performed using Oakforest-PACS at the Center for Computational Sciences, University of Tsukuba, and XC40 at Yukawa Institute for
 Theoretical Physics, Kyoto University.

\bibliography{Ti48}

\end{document}